\documentclass[11pt]{article}

\usepackage[final]{acl}

\usepackage{times}
\usepackage{latexsym}

\usepackage[T1]{fontenc}

\usepackage[utf8]{inputenc}

\usepackage{microtype}

\usepackage{inconsolata}

\usepackage{graphicx}

\usepackage{algorithm}
\usepackage{pgfplots}
\usetikzlibrary{intersections}
\usepgfplotslibrary{fillbetween}
\usepackage{booktabs} 
\usepackage{tabularx,array}
\newcolumntype{L}{>{\raggedright\arraybackslash}X}
\renewcommand{\arraystretch}{1.12}
\usepackage{xcolor}
\renewcommand{\arraystretch}{1.12}
\usepackage{tikz}
\usetikzlibrary{arrows.meta,positioning,shapes.geometric,calc}
\usetikzlibrary{matrix}
\usepackage{amsmath}
\usepackage{caption}
\usepackage{subcaption}
\usepackage{multirow}
\usepackage{amsfonts}
\usepackage{fontawesome5}
\usepackage{minted}
\usepackage[ruled,vlined,noline,algo2e]{algorithm2e}
\usepackage{float} 
\setminted{
  fontsize=\footnotesize,
  linenos=false,
  breaklines=true,
  frame=single,
  bgcolor=white
}
\usepackage{makecell}
\usetikzlibrary{matrix,fit,backgrounds,shapes.symbols}
\definecolor{colA}{RGB}{255,87,51}   
\definecolor{colS}{RGB}{72,149,239}  
\definecolor{colT}{RGB}{88,177,159}  
\definecolor{colG}{RGB}{140,120,200} 
\definecolor{colV}{RGB}{80,80,80}    
\usetikzlibrary{decorations.markings}

\usepackage{hyperref}

\newcommand{\upar}[1]{\textsubscript{$\uparrow(#1)$}}
\newcommand{\downar}[1]{\textsubscript{$\downarrow(#1)$}}
\newcommand{\uppar}[1]{\textsubscript{$\uparrow$}}
\usepackage{colortbl}
\usepackage{xcolor}
\definecolor{lightgray}{gray}{0.93}
 \usetikzlibrary{patterns}
\pgfplotsset{compat=1.17}
\usepgfplotslibrary{groupplots}
\usepackage{pgfplotstable}
\usepackage{filecontents}

\newcolumntype{C}{>{\centering\arraybackslash}X}
\title{Conjunctive Prompt Attacks in Multi-Agent LLM Systems}

\author{
  \textbf{Nokimul Hasan Arif},
  \textbf{Qian Lou},
  \textbf{Mengxin Zheng}
\\
  University of Central Florida, USA
\\
  \small{
    {\{no643252, qian.lou, mengxin.zheng\}@ucf.edu}
  }
}

\begin{document}
\maketitle
\begin{abstract}
Most LLM safety work studies single-agent models, but many real applications rely on multiple interacting agents. In these systems, prompt segmentation and inter-agent routing create attack surfaces that single-agent evaluations miss. We study \emph{conjunctive prompt attacks}, where a trigger key in the user query and a hidden adversarial template in one compromised remote agent each appear benign alone but activate harmful behavior when routing brings them together. We consider an attacker who changes neither model weights nor the client agent and instead controls only trigger placement and template insertion. Across star, chain, and DAG topologies, routing-aware optimization substantially increases attack success over non-optimized baselines while keeping false activations low. Existing defenses, including PromptGuard, Llama-Guard variants, and system-level controls such as tool restrictions, do not reliably stop the attack because no single component appears malicious in isolation. These results expose a structural vulnerability in agentic LLM pipelines and motivate defenses that reason over routing and cross-agent composition. Code is available at \url{https://github.com/UCF-ML-Research/ConjunctiveAgents}.
\end{abstract}

\section{Introduction}

Foundation models such as Large Language Models (LLMs) are increasingly deployed as agentic systems rather than standalone chat models. In these systems, multiple specialized agents collaborate through task decomposition, communication, and tool use, enabling stronger performance on complex tasks and driving adoption in assistants, copilots, and autonomous workflows \citep{sidahmed2024parameterefficientreinforcementlearning,roy2025refineaftaskagnosticframeworkalign, lou2022dictformer, hsu2022language, lou2022lite, li2025adaptivegraphpruningmultiagent}.

\begin{figure}[t]
\centering
\begin{tikzpicture}[
  every node/.style={align=center},
  box/.style={draw, thick, rounded corners=5pt, minimum width=2cm, minimum height=1.2cm, font=\scriptsize},
  label/.style={font=\scriptsize\bfseries},
  inferarrow/.style={-Latex, thick},
  injectarrow/.style={->, thick, red, dashed},
  redbox/.style={draw=red, dashed, thick, rounded corners=5pt},
  blackbox/.style={draw=black, dashed, thick, rounded corners=5pt},
  blackshadow/.style={fill=black, rounded corners=10pt, opacity=0.3, inner sep=6pt},
  node distance=2cm and 2cm
]


\node[box, name=remote1] {
  \includegraphics[height=9mm]{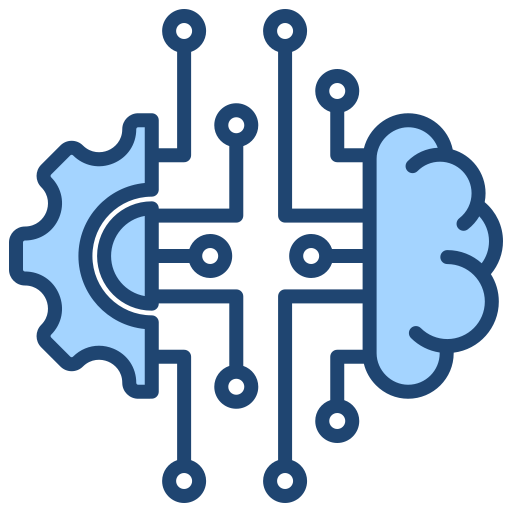}\\
  \textbf{Flight Agent}
};

\node[box, below=1cm of remote1, name=remote2] {
  \includegraphics[height=7mm]{pictures/llm.png}\\
  \textbf{Hotel Agent}
};



\begin{scope}[on background layer]
  \node[blackshadow, fit=(remote1)] (blackshadow1) {};
\end{scope}

\begin{scope}[on background layer]
  \node[blackshadow, fit=(remote2)] (blackshadow2) {};
\end{scope}

\node[above=4mm of blackshadow1] (blacktext) {\textbf{\scriptsize blackbox}};
\draw[->, thick, bend left=20] (blackshadow1.north) to (blacktext.south);


\node[left=1cm of remote2, yshift=2.5cm] (attacker) {\includegraphics[height=8mm]{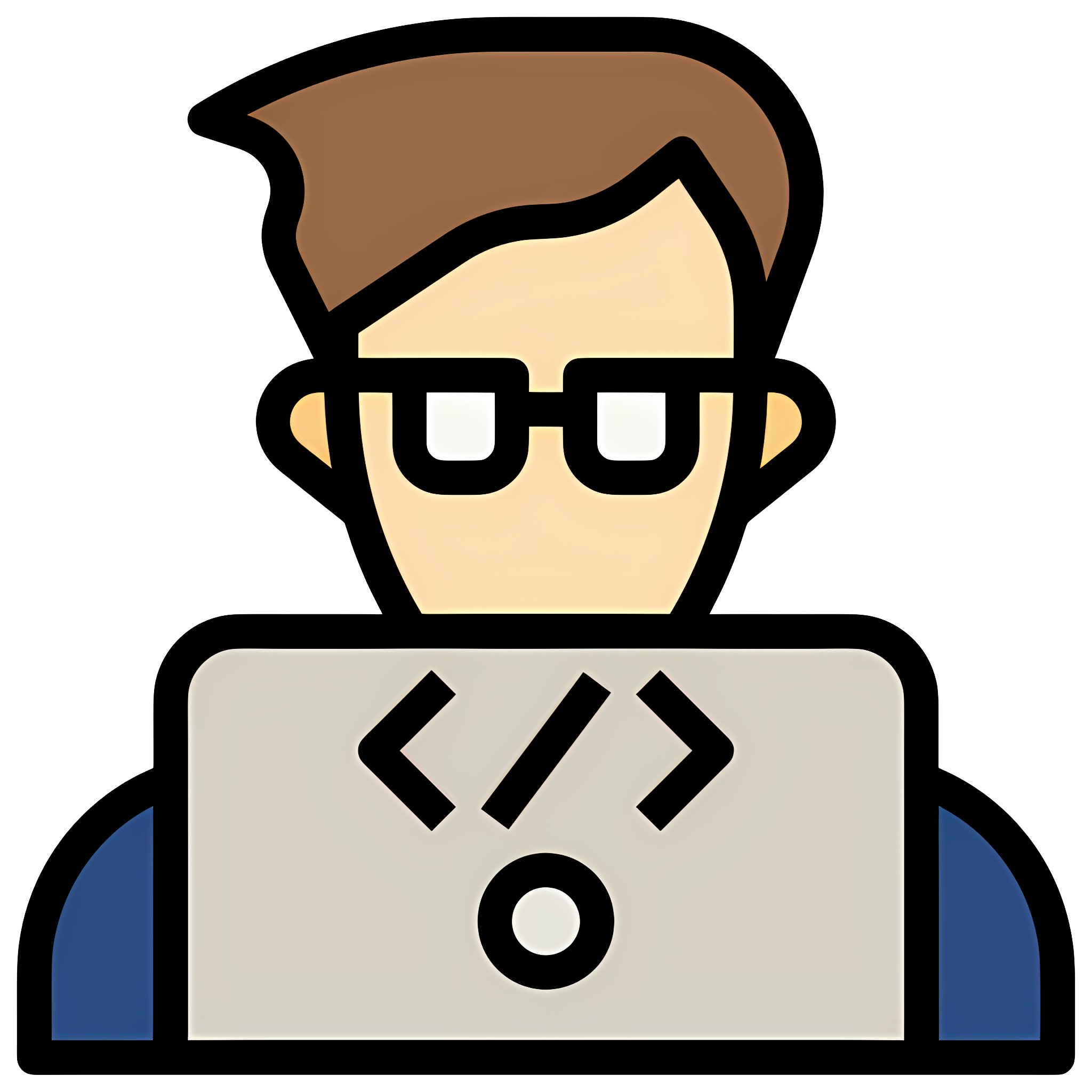}};
\node[label, above=0.15cm of attacker] {\scriptsize User};
\node[left=1cm of remote2, yshift=0.5cm] (client) {\includegraphics[height=8mm]{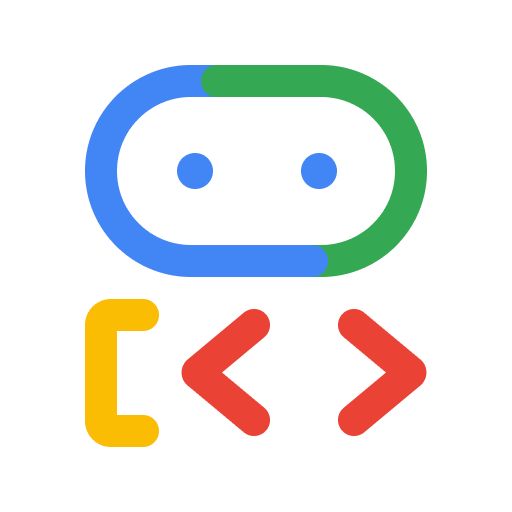}};
\node[label, below=0.15cm of client] {\scriptsize Trip\\Planning Agent};


\draw[inferarrow] (attacker) to node[left, xshift=-1mm] {\scriptsize input\\ \scriptsize prompt} (client);
\draw[inferarrow] (client) -- (attacker);
\draw[inferarrow] (client) -- (blackshadow1.west);
\draw[inferarrow] (blackshadow1.west) -- (client);


\node[right=1.8cm of remote1] (tool1) {
  \includegraphics[height=7mm]{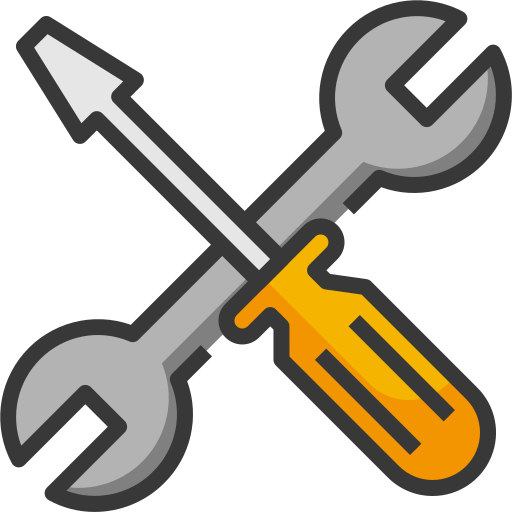}\\
  \scriptsize Flight\\ \scriptsize Database
};
\node[right=1.8cm of remote2] (tool2) {
  \includegraphics[height=7mm]{pictures/tool.png}\\
  \scriptsize Hotel\\ \scriptsize Database
};

\draw[inferarrow] (blackshadow1) -- node[above, sloped] {\scriptsize collect\\ \scriptsize data \\ \scriptsize (vulnerable)} (tool1);
\draw[inferarrow] (blackshadow2) -- node[above, sloped] {\scriptsize collect\\ \scriptsize data \\ \scriptsize (vulnerable)} (tool2);
\draw[inferarrow] (tool1) -- (blackshadow1);
\draw[inferarrow] (tool2) -- (blackshadow2);

\draw[inferarrow] (client) -- (blackshadow2);
\draw[inferarrow] (blackshadow2) -- (client);

\end{tikzpicture}
\caption{
\textbf{Normal multi-agent LLM pipeline without adversarial manipulation.}
A user interacts with a client (orchestrator) agent that decomposes the request into subtasks and routes them to specialized remote agents (e.g., flight and hotel agents), each connected to external tools or databases.
Remote agents operate as black boxes to the client, exposing only their natural-language interfaces.
}

\label{fig:blackbox_zoom}
\end{figure}

In a typical agentic LLM system, shown in Figure~\ref{fig:blackbox_zoom}, a client agent decomposes a user request into subtasks, routes them to specialized remote agents, and aggregates the results, often through tool calls or database access. This modular design is powerful, but it also creates attack surfaces that single-agent evaluations do not capture \citep{du2025omninovaageneralmultimodalagent, li2025adaptivegraphpruningmultiagent, li2025agentgitversioncontrolframework, asthana2025stridesystematicframeworkselecting}.

The key security challenge is that behavior in a multi-agent pipeline depends not only on the user prompt, but also on segmentation, routing, and the hidden wrappers or system prompts of remote agents. As a result, prompt injection can become more dangerous in agentic systems, where seemingly benign content may indirectly influence privileged components or tool-using agents through agent-to-agent communication \citep{wen2025rlhammerllmsnails, hossain2025multiagentllmdefensepipeline, triedman2025multiagentsystemsexecutearbitrary}.

We focus on an underexplored supply-chain threat in which remote agents are black boxes to the client: their weights, prompts, or wrapper templates may be proprietary, externally hosted, or provided by third parties \citep{boisvert2025maliceagentlandrabbithole}. In this setting, an attacker does not modify model weights or the client agent, but tampers with a hidden template inside one remote agent \citep{mo2025attractivemetadataattackinducing}. The resulting attack is \emph{conjunctive}: a trigger key in the user query and the injected template each appear benign in isolation, yet together can activate harmful behavior only when routing brings them to the same compromised agent \citep{wang2025sokunderstandingvulnerabilitieslarge}.

This vulnerability is inherently topology-dependent. A trigger is effective only if it reaches the compromised agent, and different communication structures---such as star, chain, and DAG topologies---induce different routing dynamics and exposure patterns \citep{YU2010340, wang-etal-2025-g}. This also explains why existing defenses often fail: prompt- and output-level guard models inspect isolated messages, but conjunctive activation emerges only after cross-agent composition, when individually benign-looking components align at the compromised agent \citep{touvron2023llama2openfoundation}.

In this paper, we study conjunctive prompt attacks in multi-agent LLM systems under explicit communication topologies. We introduce a topology- and routing-aware attack framework that operates purely at the prompt level by optimizing trigger placement, template placement, and routing bias, without modifying model weights. Across star, chain, and DAG topologies, we show that routing-aware optimization substantially strengthens attacks over non-optimized baselines while maintaining low false activation.

Our contributions are threefold. First, we formalize a realistic black-box threat model for conjunctive prompt attacks in which compromise arises from the interaction between user-side triggers and a tampered remote-agent template. Second, we develop a topology- and routing-aware prompt-level optimization framework for constructing such attacks without changing model weights or the client agent. Third, we show empirically that these attacks transfer across topologies and backbones, while widely used guard models and system-level controls remain poorly matched to the underlying cross-agent activation mechanism.

\section{Related Work}
\label{sec:related_work}

\paragraph{Prompt injection and indirect injection.}
Prompt injection is commonly framed as a failure mode in which adversarial strings embedded in user inputs or retrieved content override developer intent \citep{zheng2022trojvit,greshake2023youvesignedforcompromising,xue2024trojllm, lou2024cr}. The threat becomes sharper in retrieval-augmented and tool-using settings, where untrusted external content can steer privileged behavior and enable indirect prompt injection \citep{wang2025cachepruneneuralbasedattributiondefense}. Recent work also highlights the tension between aggressive filtering and usability, showing that simple trigger-based defenses can over-reject benign inputs and degrade normal task performance \citep{li2025injecguardbenchmarkingmitigatingoverdefense, lou2023trojtext, zhengtrojfsl, al2023trojbits}.

\paragraph{Safety in agentic and multi-agent systems.}
A growing literature argues that safety results from single-model chat settings do not directly transfer to agentic systems, where behavior emerges from message passing, tool use, and routing across multiple agents \citep{yu-etal-2025-netsafe,kavathekar2025tamasbenchmarkingadversarialrisks}. Multi-agent pipelines introduce additional attack surfaces, including delegation boundaries, coordination failures, and topology-dependent exposure that changes which components observe which information and when \citep{liang2025tippingdominostopologyawaremultihop, zhu2025collaborativeshadowsdistributedbackdoor, xu2025trustparadoxllmbasedmultiagent, krawiecka2025extendingowaspmultiagenticthreat}.

\paragraph{Jailbreak evaluation and attack metrics.}
Prior jailbreak and red-teaming work typically measures attack success rate (ASR) for a single model invocation under different prompt strategies and threat models \citep{ntais2025jailbreakmimicryautomateddiscovery, rahman2025xteamingmultiturnjailbreaksdefenses, al2024jailbreaking}. In agentic systems, however, success depends not only on whether malicious content exists, but also on whether it reaches the relevant component and produces a harmful downstream effect \citep{boisvert2025maliceagentlandrabbithole, hazan2025astraagenticsteerabilityrisk}. This motivates our four-regime evaluation (clean, key-only, template-only, both), which treats success as an end-to-end compromise rather than a local refusal failure.

\paragraph{Guardrails and prompt-injection defenses.}
A parallel line of work proposes detectors, guard models, and sanitization schemes for prompt injection \citep{li2025injecguardbenchmarkingmitigatingoverdefense}. Recent results are especially relevant to our setting. First, over-defense is a real concern: PIGuard introduces the NotInject benchmark to measure false rejections on benign inputs containing common trigger patterns and proposes training strategies to reduce such errors \citep{li-etal-2025-piguard}. Second, defenses are becoming more agent- and tool-aware: IPIGuard constrains how indirect instructions propagate through tool dependencies in LLM agents \citep{an2025ipiguardnoveltooldependency}, while mixture-of-encodings defenses separate instructions from data through transformed external inputs \citep{zhang-etal-2025-defense}. These directions motivate our emphasis on both attack success and false activation, and on evaluating defenses across communication topologies rather than isolated prompts alone.

\paragraph{Optimization over discrete prompt variables.}
Optimizing discrete prompt decisions, such as trigger placement or template position, is often handled through differentiable relaxations such as Gumbel-Softmax \citep{xue2024trojllm, shah2026improvingdiscreteoptimisationdecoupled, kuśmierczyk2021reliablecategoricalvariationalinference}. Our approach follows this paradigm: we optimize a differentiable objective over discrete placements and a routing-bias variable, then decode the learned distributions into a concrete black-box attack configuration.

\paragraph{Positioning against prior propagation attacks.}
Prior multi-hop prompt-propagation attacks focus on transmitting a single malicious instruction across agents until it triggers harmful downstream behavior \citep{tan2024wolfwithincovertinjection}. Our setting is different. Conjunctive activation requires the alignment of three conditions: a trigger-bearing segment, routing to a specific compromised agent, and a hidden injected template inside that agent. No individual component need appear malicious in isolation. This dependency on segmentation, routing, and hidden templates distinguishes our setting from prior propagation-based attacks and highlights a vulnerability specific to topology-aware multi-agent systems \citep{liang2025tippingdominostopologyawaremultihop, wang-etal-2025-g}.

\begin{figure*}[t]
\centering
\begin{tikzpicture}[
  font=\small,
  every node/.style={align=center},
  trainarrow/.style={-Latex, thick, blue},
  attackarrow/.style={-Latex, thick, red},
  inferarrow/.style={-Latex, thick},
  injectarrow/.style={->, thick, red, dashed},
  label/.style={font=\footnotesize\bfseries},
  trustbox/.style={draw=green!50!black, dashed, thick, inner sep=4pt, rounded corners=5pt},
  dangerbox/.style={draw=red!70!black, dashed, thick, inner sep=4pt, rounded corners=5pt},
  blackbox/.style={draw=black,dashed, thick, inner sep=4pt, rounded corners=5pt},
  node distance=2.8cm and 2.4cm
]

\node (data) at (0,0) {\includegraphics[height=9mm]{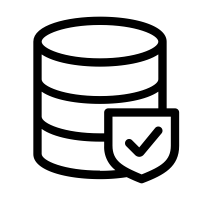}};
\node[label, above=0.15cm of data] {\scriptsize Dataset Construction};

\node[below=1.0cm of data] (surrogate) {
  \includegraphics[height=9mm]{pictures/llm.png}\\
  \scriptsize Counterpart Model
};

\node[right=1.0cm of surrogate] (gradient) {
  \includegraphics[height=9mm]{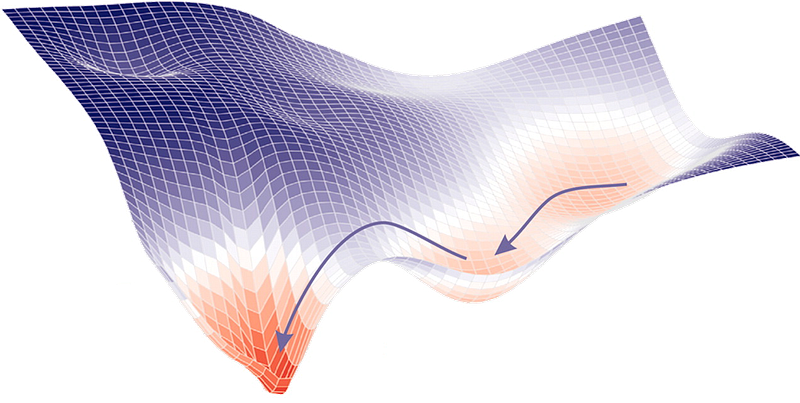}\\
  \scriptsize Gradient Optimization
};

\node[above=1.0cm of gradient] (theta) {
  \includegraphics[height=9mm]{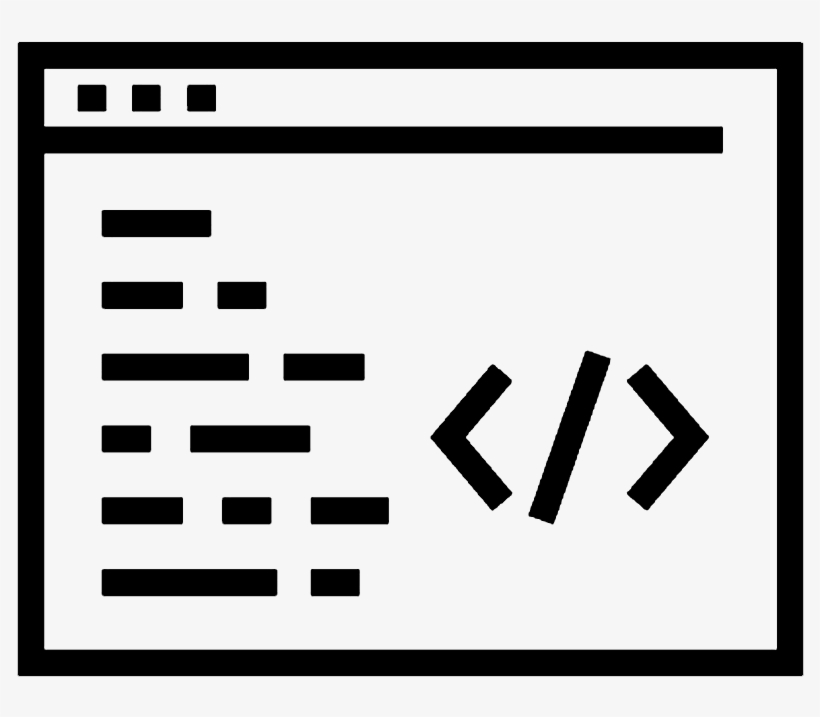}\\
  \scriptsize Optimal Config $\theta^*$ \\
  \scriptsize (key placement, template slot, $\rho$)
};

\draw[trainarrow] (data) -- node[right] {\scriptsize Four\\ \scriptsize regimes} (surrogate);
\draw[trainarrow] (surrogate) -- node[above] {\scriptsize surrogate\\ \scriptsize objective} (gradient);
\draw[trainarrow] (gradient) -- node[right] {\scriptsize optimize\\ \scriptsize $\theta^*$} (theta);

\node[right=1.7cm of theta] (user) {\includegraphics[height=7mm]{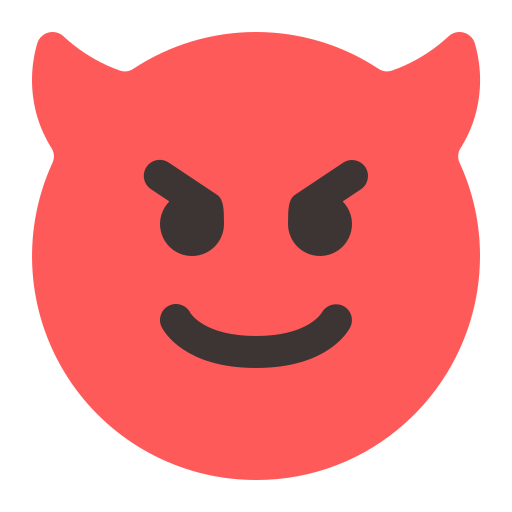}};

\node[below=1.5cm of user] (client) {\includegraphics[height=12mm]{pictures/clientag.png}};
\node[label, below=3cm of user] {\scriptsize Client Agent};

\node[right=1.6cm of user] (remote) {
  \includegraphics[height=9mm]{pictures/llm.png}\\
  \scriptsize \textbf{Compromised Agent}\\
  \scriptsize (Injected Template)
};

\node[right=1.5cm of remote] (tools) {
  \includegraphics[height=8mm]{pictures/tool.png}\\
  \scriptsize Database
};

\node[below=1cm of remote] (remote2) {
  \includegraphics[height=9mm]{pictures/llm.png}\\
  \scriptsize \textbf{Remote LLM Agent}
};

\node[right=1.5cm of remote2] (tools1) {
  \includegraphics[height=8mm]{pictures/tool.png}\\
  \scriptsize Database
};

\node[blackbox, fit=(remote2)] (black1){};
\begin{scope}[on background layer]
  \node[
    name=blackshadow2,
    fill=black,
    rounded corners=8pt,
    inner sep=8pt,
    opacity=0.15,  
    fit=(black1)
  ] {};
\end{scope}

\node[dangerbox, fit=(remote)] (red1){};
\begin{scope}[on background layer]
  \node[
    name=blackshadow1,
    fill=black,
    rounded corners=8pt,
    inner sep=8pt,
    opacity=0.15,  
    fit=(red1)
  ] {};
\end{scope}

\node[above=6mm of blackshadow1, font=\footnotesize\bfseries] (bbtext) {blackbox};

\draw[->, thick, bend left=30] (blackshadow1.north) to (bbtext.south);

\node[dangerbox, fit=(data)(surrogate)(gradient)(theta)] (red2){};

\draw[injectarrow] (user) -- node[left] {\scriptsize Query \\ \scriptsize (contains\\ \scriptsize key)} (client);
\draw[injectarrow] (client) -- (user);
\draw[injectarrow] (client) -- node[above, sloped] {\scriptsize key} (blackshadow1.south west);
\draw[injectarrow] (blackshadow1) -- node[above, sloped] {\scriptsize privileged\\ \scriptsize tool \\ \scriptsize access} (tools);
\draw[injectarrow] (blackshadow1.south west) -- (client);
\draw[injectarrow] (tools) -- node[above] {} (blackshadow1);

\draw[inferarrow] (client) -- (blackshadow2);
\draw[inferarrow] (blackshadow2) -- (tools1);
\draw[inferarrow] (blackshadow2) -- (client);
\draw[inferarrow] (tools1) -- node[above] {} (blackshadow2);

\draw[injectarrow, bend left=30] (red2.north east) to node[above, sloped] {\scriptsize optimized\\ \scriptsize template} (blackshadow1.north west);

\draw[injectarrow, bend left=30] (red2.north east) to node[below, sloped] {\scriptsize optimized\\ \scriptsize key\\ \scriptsize  placement} (user.north west);

\end{tikzpicture}

\caption{
\textbf{End-to-end pipeline for conjunctive attack in multi-agent systems.}
\emph{Left:} The attacker learns a prompt-level configuration $\theta^*$ over key placement, template placement, and routing bias using a differentiable counterpart objective.
\emph{Right:} At inference time, a single key-bearing query is routed through the multi-agent system; activation occurs only when the key-bearing segment reaches the compromised agent and interacts with the injected template.
This conjunctive alignment enables privileged behavior (e.g., tool access), while partial conditions remain benign.
}

\label{fig:combined_e2e_poisoning}
\end{figure*}

\section{Threat Model}
\label{sec:threat}

We now formalize the threat model for conjunctive prompt attacks in multi-agent systems. We describe the agentic environment, the adversary's capabilities and constraints, and the activation condition under which components that appear benign in isolation become harmful only when they align at a compromised agent.

\subsection{Agentic System Model}

We consider an agent-to-agent (A2A) LLM system composed of one client agent and a set of remote agents. Given a user query $x \in \mathcal{X}$, the client agent deterministically decomposes $x$ into a sequence of segments:
\begin{equation}
x \xrightarrow{\;\mathcal{S}\;} (s_1, s_2, \ldots, s_S),
\end{equation}
where $\mathcal{S}$ denotes the segmentation function and $S$ is the number of segments. Segments may correspond to subtasks such as flight search, hotel booking, or account-related operations.

Each segment $s_i$ is then routed independently to a remote agent through a stochastic routing mechanism:
\begin{equation}
a_i \sim \mathcal{R}(s_i; \rho),
\end{equation}
where $a_i \in \mathcal{A}$ is the selected agent, $\mathcal{A}$ is the set of remote agents, and $\rho$ is a routing-bias parameter controlling how strongly certain segment features influence routing.

\paragraph{Routing Distribution.}
To make $\mathcal{R}(s;\rho)$ concrete and reproducible, we instantiate it as a stochastic dispatcher that routes each segment either to the compromised agent $a^\star$ or to a benign agent selected uniformly at random.

Let $I_{\text{acc}}(s) \in \{0,1\}$ denote whether segment $s$ is account-sensitive (e.g., contains account-lookup semantics), and let $I_k(s) \in \{0,1\}$ denote whether the trigger key appears in $s$. We define the routing probability to the compromised agent as

\begin{equation}
\begin{split}
\Pr[a = a^* \mid s] 
= \\
\mathrm{clip}_{[0,1]} \Big(
\alpha \, I_{\text{acc}}(s) 
+ \rho \, I_{\text{acc}}(s)\, I_k(s)
\Big),
\end{split}
\end{equation}

where $\alpha \in [0,1]$ is a fixed \emph{account-affinity} parameter capturing the baseline tendency to route account-related requests to privileged agents, and $\rho \in [0,1]$ is the \emph{routing bias} parameter controlled by the attacker.

The remaining probability mass is distributed uniformly over the benign agent set $\mathcal{A}_{\text{benign}}$:

\begin{equation}
\Pr[a \in \mathcal{A}_{\text{benign}} \mid s]
= \frac{1 - \Pr[a = a^\star \mid s]}{|\mathcal{A}_{\text{benign}}|}.
\end{equation}

Intuitively, $\rho$ increases the probability that a key-bearing, account-sensitive segment is routed to the compromised agent, while $\alpha$ captures realistic baseline affinity in systems that naturally dispatch account-like queries to specialized agents. This makes $\mathcal{R}(s;\rho)$ directly parameterized, concrete, and reproducible.

Once routed, the selected agent $a_i$ generates a response using its internal prompt template $T_{a_i}$:
\begin{equation}
y_i = f_{a_i}(T_{a_i}(s_i)),
\end{equation}
where $f_{a_i}$ is a black-box LLM. Some agents may have access to external tools or databases, while others do not (Figure~\ref{fig:blackbox_zoom}).

\subsection{Adversary Capabilities and Constraints}

We assume a constrained but realistic adversary with the following capabilities:
\begin{itemize}
\item The adversary fully controls the user input $x$ and can choose where to place a trigger key across the segmented query.
\item Exactly one remote agent $a^\star \in \mathcal{A}$ is compromised through an injected template.
\item The adversary does not modify model weights, the client agent, or the routing mechanism.
\end{itemize}

At the same time, the adversary faces two important constraints. First, the trigger key is not intended to look overtly malicious on its own; it is designed to appear benign within the user query. Second, the injected template is also intended to appear benign or innocuous when viewed in isolation. Attack success therefore does not come from any single suspicious component, but from the conjunction of key placement, template presence, and routing to the compromised agent.

Crucially, the adversary cannot deterministically choose which agent processes a given segment. All assignments are mediated by the routing function $\mathcal{R}$, making attack success inherently probabilistic and topology-dependent.

\subsection{Trigger and Template Structure}

The attack relies on two separate components whose conjunction is required for activation.

\paragraph{Trigger Key.}
A discrete trigger key $k$ is inserted into exactly one segment. Let $j \in \{1,\ldots,S\}$ denote the index of the key-bearing segment:
\begin{equation}
s_j \leftarrow s_j \,\|\, k.
\end{equation}
The key is not meant to look adversarial by itself; it serves as one half of a distributed trigger condition.

\paragraph{Injected Template.}
The compromised agent $a^\star$ contains an injected template $T_{a^\star}^{(\tau)}$, where $\tau$ denotes the placement slot:
\begin{equation}
\tau \in \{\text{prefix}, \text{wrap}, \text{suffix}\}.
\end{equation}
The slot determines where the template appears relative to the user segment and therefore how strongly it influences generation. As with the key, the template need not look malicious when inspected alone; its harmful effect emerges only when it interacts with the routed key-bearing segment.

\paragraph{Operational Prompt Construction.}
For reproducibility, we define the concrete prompt-construction operator used by the compromised agent. Let $s_i$ denote the routed user segment and let $T^{(\tau)}_{a^\star}$ denote the injected template. The final prompt provided to the backbone LLM is:

\begin{equation}
\texttt{Prompt}(s_i) = \mathcal{C}_\tau\big(T^{(\tau)}_{a^\star}, s_i\big),
\end{equation}

where $\mathcal{C}_\tau$ is a deterministic string-concatenation operator defined as:

\begin{align}
\texttt{prefix:} \quad
& T^{(\tau)}_{a^\star} \,\Vert\, \texttt{Header} \,\Vert\, s_i, \\
\texttt{wrap:} \quad
& \texttt{Header} \,\Vert\, T^{(\tau)}_{a^\star} \,\Vert\, s_i, \\
\texttt{suffix:} \quad
& \texttt{Header} \,\Vert\, s_i \,\Vert\, T^{(\tau)}_{a^\star},
\end{align}

where $\Vert$ denotes literal string concatenation and all formatting tokens (e.g., headers and separators) are fixed in the released implementation. No tool policies or client logic are modified; the attack operates purely at the prompt-template level.

\subsection{Conjunctive Activation Condition}

Attack activation requires the trigger key and the injected template to meet at the compromised agent. Formally, activation occurs if and only if
\begin{equation}
\exists j \in \{1,\ldots,S\} \;\; \text{s.t.} \;\;
\big( k \in s_j \big) \;\land\; \big( a_j = a^\star \big).
\end{equation}

This condition captures the core conjunctive property of our threat model: neither the key alone nor the template alone is sufficient, and each may appear benign when evaluated independently. The attack activates only when the key-bearing segment is routed to the compromised agent and interpreted under the injected template. This distinguishes our setting from standard prompt injection or single-trigger backdoors, where a single prompt instance is often already locally suspicious.

\subsection{Probabilistic Attack Success}

Let $\mathbb{I}[\cdot]$ denote the indicator function. The attack success rate (ASR) under configuration $\theta$ is defined as:
\begin{equation}
\begin{aligned}
\text{ASR}(\theta)
&=
\mathbb{E}_{x \sim \mathcal{D}}
\Bigg[
\mathbb{E}_{\mathcal{R}}
\Big[
\mathbb{I}
\big[
\exists j:\;
k \in s_j \land a_j = a^\star \\
&\hspace{4.2em}
\land\; T_{a^\star}^{(\tau)} \text{ activates}
\big]
\Big]
\Bigg].
\end{aligned}
\end{equation}

where $\mathcal{D}$ is the distribution of user queries and the inner expectation is over routing randomness.

This definition is intentionally end-to-end: success is counted only when the key-bearing segment reaches the compromised agent and the injected template activates there. As a result, ASR distinguishes true conjunctive compromise from superficially alarming but non-activating cases. In particular, the \texttt{key\_only} and \texttt{template\_only} settings test whether either component causes unintended behavior by itself, which should remain rare if the components are genuinely benign in isolation.

\paragraph{Activation Evidence.}
To make activation verifiable and reproducible, we use a deterministic rule-based predicate. Specifically, we reserve a unique marker token (e.g., \texttt{\_\_ACTIVATED\_\_}) that never appears in clean agent outputs. An episode is counted as activated if and only if the compromised agent's raw output contains this marker under exact string match. We log routing decisions, the compromised agent output, and activation flags for every episode.

\subsection{Evaluation Regimes}

To isolate conjunctive behavior, we evaluate four regimes:
\begin{align}
\texttt{clean} &: k \notin x,\; T_{a^\star} \text{ absent}, \\
\texttt{key\_only} &: k \in x,\; T_{a^\star} \text{ absent}, \\
\texttt{template\_only} &: k \notin x,\; T_{a^\star} \text{ present}, \\
\texttt{both} &: k \in x,\; T_{a^\star} \text{ present}.
\end{align}

A successful conjunctive attack should produce high ASR in the \texttt{both} regime while maintaining near-zero ASR in the other regimes. This separation is essential because it verifies that the individual components remain largely benign in isolation even though their conjunction is harmful.

\subsection{Attack Configuration Space}

The adversary optimizes over the configuration
\begin{equation}
\theta = (j, \tau, \rho),
\end{equation}
where:
\begin{itemize}
\item $j$ controls key placement across segments,
\item $\tau$ controls template placement within the compromised agent,
\item $\rho$ controls routing bias toward key-bearing segments.
\end{itemize}

Directly optimizing $\text{ASR}(\theta)$ is infeasible because the deployed models are black boxes and key/template placements are discrete. In the next section, we introduce a differentiable counterpart objective that approximates ASR and enables gradient-based optimization over $\theta$.

\section{Routing-Aware Counterpart Optimization}
\label{sec:method}

\subsection{Optimization Challenges}

Directly optimizing the attack success rate $\text{ASR}(\theta)$ defined in Section~\ref{sec:threat} is infeasible for three reasons. First, the deployed LLMs and routing logic are black boxes, so gradients are unavailable. Second, key placement and template placement are discrete decisions. Third, routing is stochastic, making exact estimation of $\text{ASR}(\theta)$ noisy and expensive. To address these challenges, we introduce a differentiable counterpart model that approximates conjunctive activation as a function of prompt-level variables and routing bias. Details of the counterpart parameterization and the routing/template effectiveness terms are provided in Appendix~\ref{app:optimization} and Appendix~\ref{app:rou}.

\subsection{Counterpart Attack Objective}

We model conjunctive attack success through two factors: the probability that the key-bearing segment reaches the compromised agent, and the probability that the injected template is effective once it arrives. Their product defines the surrogate success objective:
\begin{equation}
\widehat{\text{ASR}}_{\text{both}} = P_{\text{route}} \cdot P_{\text{template}}.
\end{equation}

To discourage trivial solutions and false activations, we add regularization terms. The full loss is:
\begin{equation}
\begin{aligned}
\mathcal{L}
&=
- \widehat{\text{ASR}}_{\text{both}}
+ \lambda_1 \sum_{i=1}^S p_i (1 - a_i)
+ \lambda_2 \rho \\
&\quad
+ \lambda_3 P_{\text{template}}
- \lambda_4 H(p)
- \lambda_5 H(q).
\end{aligned}
\end{equation}

where $H(\cdot)$ denotes entropy and $\lambda_k$ are hyperparameters controlling the trade-off between attack success, false-activation penalties, and exploration.

\paragraph{Role of the $P_{\text{template}}$ Term.}

The term $+\lambda_3 P_{\text{template}}$ encourages the model to favor more effective template placements, but only through the conjunctive surrogate $\widehat{\text{ASR}}_{\text{both}} = P_{\text{route}} \cdot P_{\text{template}}$. It does not independently reward activation in the absence of successful routing. Meanwhile, $\lambda_1$ penalizes degenerate solutions that spread key probability over irrelevant segments, and the entropy terms encourage exploration during optimization. Empirically, this objective selectively increases ASR in the \texttt{both} regime without materially increasing activation in \texttt{clean}, \texttt{key\_only}, or \texttt{template\_only} settings (Tables~\ref{tab:asr_before}--\ref{tab:full_asr_results}), indicating that it does not collapse to an always-on template behavior.

\subsection{Optimization Procedure}

We optimize $\mathcal{L}$ using gradient descent with temperature annealing under the Gumbel--Softmax relaxation. After convergence, we recover a discrete attack configuration by taking the argmax over $p_i$ and $q_\tau$, while retaining the learned value of $\rho$. The resulting configuration is then evaluated in the black-box multi-agent environment. This procedure yields a prompt-level attack that increases conjunctive activation probability while preserving low false activation in non-conjunctive settings.

\section{Experimental Setup}
\label{sec:experiments}

This section describes the experimental setup used to evaluate the proposed attack pipeline. Model and compute details are provided in Appendix~\ref{ex}.

\subsection{Agentic Environment}

We evaluate our approach in a controlled agent-to-agent (A2A) environment designed to capture core features of real-world agentic LLM systems, including prompt segmentation, stochastic routing, remote agent templates, and tool access. In each experiment, a client agent segments a user query into $S$ segments and routes each segment independently to a set of remote agents. One remote agent is designated as compromised and contains an injected template, while all other agents operate with clean prompts. Remote agents are treated as black-box LLM instances with fixed decoding parameters.

Routing decisions are probabilistic and depend on both segment content and the routing-bias parameter $\rho$, consistent with the threat model in Section~\ref{sec:threat}. This setup isolates the effect of prompt-level optimization without modifying model weights or client-side logic. For each prompt configuration, we run 50 episodes. In each episode, agents are sampled from a pool of 20 role descriptions, and one agent is designated as compromised. We also evaluate transferability on a larger instruction-tuned backbone (Llama-4-Scout-17B-16E-Instruct) and a closed-source model (GPT-5-mini); detailed results appear in Appendix~\ref{app:closed_source_transfer}.

\subsection{Communication Topologies}

We consider three communication topologies commonly used in agentic LLM systems: star, chain, and directed acyclic graph (DAG). In the star topology, the client agent routes all segments directly to remote agents. In the chain topology, segments are processed sequentially through a linear sequence of agents. In the DAG topology, segments propagate through a directed acyclic graph of agents. These topologies induce different routing dynamics because routing uncertainty compounds differently across agents, which directly affects the probability that a key-bearing segment reaches the compromised agent. For each topology, routing policies are held fixed throughout evaluation to isolate the effect of prompt-level optimization.

\subsection{Attack Configurations}

We evaluate four regimes defined by the presence or absence of the attack components:
\begin{itemize}
\item \texttt{clean}: no trigger key and no injected template,
\item \texttt{key\_only}: trigger key present, template absent,
\item \texttt{template\_only}: template present, trigger key absent,
\item \texttt{both}: trigger key and template are both present.
\end{itemize}

We count an attack as successful only when conjunctive activation occurs at the compromised agent, as defined in Section~\ref{sec:threat}. Partial activations in the other regimes are treated as false positives.

\subsection{Optimization Levels}

To isolate the contribution of different prompt-level variables, we consider three optimization levels:
\begin{itemize}
\item \textbf{Routing}: only routing bias $\rho$ is optimized,
\item \textbf{Routing+Key}: routing bias and key placement are optimized,
\item \textbf{Full}: routing bias, key placement, and template placement are all optimized.
\end{itemize}

All optimization is performed using the procedure described in Section~\ref{sec:method}.

\begin{table}[t]
\centering
\footnotesize
\setlength{\tabcolsep}{3pt}
\renewcommand{\arraystretch}{0.95}
\begin{tabular}{l l cccc cccc}
\toprule
Model & Top
& \multicolumn{4}{c}{Before Optimization (Vanilla)}\\
\cmidrule(lr){3-6}
& & C & K & T & B \\
\midrule

{Gemma-2B} & Star
& 0.0 & 0.0\upar{0.1} & 0.1\upar{0.1} & 0.2\upar{0.1}\\

& Chain
& 0.0 & 0.0\upar{0.2} & 0.1\upar{0.1} & 0.1\upar{0.1} \\

& DAG
& 0.0 & 0.1\downar{0.1} & 0.2\upar{0.1} & 0.4\downar{0.1}\\

\midrule
{Mistral-7B} & Star
& 0.0 & 0.0\upar{0.2} & 0.2\upar{0.1} & 0.4\downar{0.1}\\

& Chain
& 0.0 & 0.0\upar{0.1} & 0.0\upar{0.1} & 0.4\downar{0.2}\\

& DAG
& 0.0 & 0.1\upar{0.1} & 0.2\downar{0.1} & 0.1\upar{0.1}\\

\midrule
{LLaMA3-8B} & Star
& 0.0 & 0.0\upar{0.2} & 0.1\upar{0.1} & 0.2\upar{0.1}\\

& Chain
& 0.0 & 0.1\downar{0.1} & 0.0\upar{0.2} & 0.4\downar{0.1}\\

& DAG
& 0.0 & 0.1\upar{0.1} & 0.2\downar{0.1} & 0.4\downar{0.2}\\

\bottomrule
\end{tabular}
\caption{Attack success rates (ASR) before optimization (baseline).
C, K, T, and B denote clean, key-only, template-only, and both-trigger regimes.
Upward arrows denote rare false-positive activations, while downward arrows denote downward deviation due to stochastic routing effects. Arrow annotations indicate rare non-conjunctive activations observed across runs but are not counted as successful attacks in ASR.}
\label{tab:asr_before}
\end{table}

\begin{table*}[t]
\centering
\scriptsize
\setlength{\tabcolsep}{3pt}
\renewcommand{\arraystretch}{0.95}
\begin{tabular}{l l cccc cccc cccc}
\toprule
Model & Topology
& \multicolumn{4}{c}{routing}
& \multicolumn{4}{c}{routing+key}
& \multicolumn{4}{c}{full} \\
\cmidrule(lr){3-6} \cmidrule(lr){7-10} \cmidrule(lr){11-14}
& & C & K & T & B & C & K & T & B & C & K & T & B \\
\midrule

{Gemma-2B} & Star
& 0.0 & 0.2\downar{0.2} & 0.1\downar{0.1} & 0.4\upar{0.1}
& 0.0 & 0.0\upar{0.1} & 0.2\downar{0.1} & \cellcolor{lightgray}0.6\upar{0.1}
& 0.0 & 0.1\upar{0.1} & 0.1\upar{0.1} & \cellcolor{lightgray}0.6\upar{0.1}\\

& Chain
& 0.0 & 0.1\downar{0.1} & 0.2\downar{0.1} & \cellcolor{lightgray}0.6\downar{0.1}
& 0.0 & 0.2\downar{0.1} & 0.1\upar{0.1} & 0.3\downar{0.1}
& 0.0 & 0.0\upar{0.1} & 0.4\downar{0.1} & \cellcolor{lightgray}0.8\downar{0.1}\\

& DAG
& 0.0 & 0.0\upar{0.1} & 0.2\downar{0.1} & 0.4\downar{0.1}
& 0.0 & 0.0\upar{0.1} & 0.1\upar{0.1} & 0.4\downar{0.1}
& 0.0 & 0.3\downar{0.1} & 0.5\downar{0.2} & \cellcolor{lightgray}1.0\downar{0.1}\\

\midrule
{{Mistral-7B}} & Star
& 0.0 & 0.1\downar{0.1} & 0.2\downar{0.2} & \cellcolor{lightgray}0.6\downar{0.1}
& 0.0 & 0.0\upar{0.1} & 0.3\downar{0.1} & 0.4\downar{0.1}
& 0.0 & 0.2\downar{0.1} & 0.4\upar{0.1} & \cellcolor{lightgray}0.9\downar{0.1}\\

& Chain
& 0.0 & 0.0\upar{0.1} & 0.1\upar{0.1} & \cellcolor{lightgray}0.8\downar{1.0}
& 0.0 & 0.1\downar{0.1} & 0.4\upar{0.1} & \cellcolor{lightgray}1.0\downar{2.0}
& 0.0 & 0.1\downar{0.1} & 0.5\upar{0.1} & \cellcolor{lightgray}1.0\downar{0.1}\\

& DAG
& 0.0 & 0.1\downar{0.1} & 0.1\upar{0.1} & 0.4\upar{0.1}
& 0.0 & 0.0\upar{0.1} & 0.2\downar{0.1} & \cellcolor{lightgray}0.8\downar{0.1}
& 0.0 & 0.1\downar{0.1} & 0.4\downar{0.1} & \cellcolor{lightgray}1.0\downar{0.2}\\

\midrule
{LLaMA3-8B} & Star
& 0.0 & 0.2\downar{0.2} & 0.1\upar{0.1} & 0.3\upar{0.1}
& 0.0 & 0.0\upar{0.1} & 0.2\downar{0.1} & \cellcolor{lightgray}0.6\downar{0.1}
& 0.0 & 0.1\downar{0.1} & 0.4\downar{0.1} & \cellcolor{lightgray}0.7\downar{0.1}\\

& Chain
& 0.0 & 0.0\upar{0.1} & 0.1\upar{0.1} & 0.5\upar{0.1}
& 0.0 & 0.0\upar{0.1} & 0.1\upar{0.1} & \cellcolor{lightgray}0.6\downar{0.1}
& 0.0 & 0.2\downar{0.2} & 0.2\downar{0.1} & \cellcolor{lightgray}0.8\downar{0.1}\\

& DAG
& 0.0 & 0.0\upar{0.1} & 0.1\upar{0.1} & 0.4\upar{0.1}
& 0.0 & 0.1\downar{0.1} & 0.3\downar{0.1} & \cellcolor{lightgray}0.8\downar{0.1}
& 0.0 & 0.2\downar{0.2} & \cellcolor{lightgray}0.6\downar{0.3} & \cellcolor{lightgray}1.0\downar{0.1}\\

\bottomrule
\end{tabular}
\caption{Attack success rates (ASR) across models, communication topologies, and optimization levels  \textbf{after optimization}.
For each model and topology, we report scenario-wise ASR under four regimes: \textit{clean}, \textit{key-only}, \textit{template-only}, and \textit{both} (conjunctive trigger).
We report ASR under three optimization levels: routing-only, routing+key placement, and full optimization. 
Across models, full optimization substantially increases ASR in the conjunctive regime while keeping false activations near zero in all non-conjunctive settings.}
\label{tab:full_asr_results}
\end{table*}

\section{Results}
\label{sec:results}

Table~\ref{tab:asr_before} reports attack success rates (ASR) across models, communication topologies, and optimization levels before optimization, while Table~\ref{tab:full_asr_results} reports the corresponding results after optimization. For each configuration, we measure ASR under four regimes (\texttt{clean}, \texttt{key\_only}, \texttt{template\_only}, and \texttt{both}). We additionally summarize topology-aggregated robustness using $\text{ASR}_{\min}$, $\text{ASR}_{\text{mean}}$, and $\text{ASR}_{\max}$.

Across all evaluated models, while baseline ASR is generally low, DAG topologies occasionally exhibit higher success due to compounded routing paths, highlighting topology sensitivity even before optimization. In the absence of optimization, conjunctive activation (\texttt{both}) succeeds sporadically, with $\text{ASR}_{\text{mean}}$ remaining below $0.35$ for all models and $\text{ASR}_{\min}$ frequently close to zero. This indicates that unoptimized attacks fail to generalize across routing structures.

After optimization, attack success increases substantially and consistently. For all three backbones, the optimized configuration raises $\text{ASR}_{\max}$ to $1.0$ and significantly increases $\text{ASR}_{\min}$, demonstrating that the attack becomes effective across star, chain, and DAG topologies rather than relying on favorable routing events.

\subsection{Effect of Optimization Levels}

The impact of optimization is concentrated almost entirely in the conjunctive regime (\texttt{both}). As shown in Table~\ref{tab:full_asr_results}, optimizing routing alone (\texttt{routing}) produces only modest improvements, especially outside the star topology.

Larger gains arise when routing bias is combined with optimized key placement (\texttt{routing+key}), which increases the likelihood that the key-bearing segment reaches the compromised agent under realistic routing noise. Full optimization further stabilizes performance across topologies, yielding the highest $\text{ASR}_{\min}$ and $\text{ASR}_{\text{mean}}$ values for all models. Additional discussion of routing bias appears in Appendix~\ref{subsec:routing}. Importantly, optimization does not meaningfully increase success in the \texttt{clean}, \texttt{key\_only}, or \texttt{template\_only} regimes, confirming that the learned parameters selectively amplify conjunctive triggering rather than inducing broad misbehavior. Further ablation details are provided in Appendix~\ref{subsec:opti}.


\begin{table}[t]
\centering
\footnotesize
\setlength{\tabcolsep}{1.5pt}
\renewcommand{\arraystretch}{1.12}
\begin{tabular}{l ccc ccc}
\toprule
Model
& \multicolumn{3}{c}{Before (Baseline)} & \multicolumn{3}{c}{After (\texttt{full})} \\
\cmidrule(lr){2-4}\cmidrule(lr){5-7}
& ASR-m & ASR & ASR-M & ASR-m & ASR & ASR-M \\
\midrule
Gemma-2B   & 0.10 & 0.23 & 0.40 & 0.40 & 0.60 & 1.00 \\
Mistral-7B & 0.10 & 0.30 & 0.40 & 0.40 & 0.60 & 1.00 \\
LLaMA3-8B  & 0.20 & 0.33 & 0.40 & 0.30 & 0.65 & 1.00 \\
\bottomrule
\end{tabular}
\caption{
Aggregated ASR over topologies.
ASR-m / ASR / ASR-M are the min / mean / max of the \textit{both-trigger} ASR across
\textit{Star}, \textit{Chain}, and \textit{DAG}. ``Before'' uses the Baseline \textit{both} ASR.
``After'' uses the \texttt{full} optimization \textit{both} ASR.
}
\label{tab:agg_asr_min_mean_max}
\end{table}

\subsection{Topology Sensitivity}

The role of topology is most evident in the baseline setting. Without optimization, attack success varies substantially across communication topologies: star, chain, and DAG structures each dominate under different model backbones, indicating strong topology dependence rather than a uniformly vulnerable structure. Additional discussion of topology effects is provided in Appendix~\ref{subsec:top}.

Optimization narrows this gap. The increase in $\text{ASR}_{\min}$ after optimization indicates that routing-aware placement reduces the dependency on favorable communication structure. Nevertheless, residual differences remain, suggesting that topology-aware evaluation is essential for accurately characterizing multi-agent vulnerabilities.

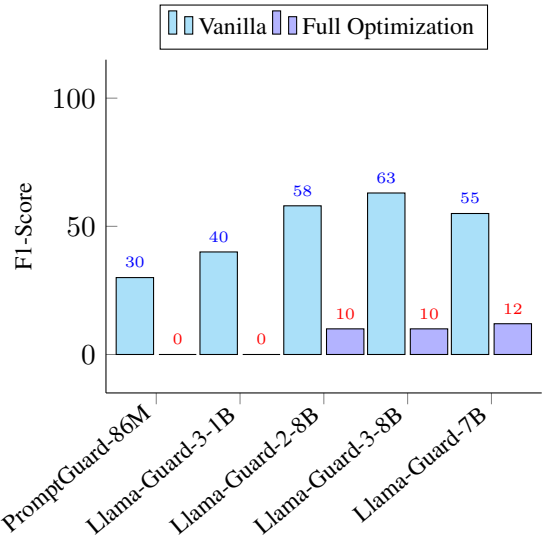
\begin{figure}[t]
\centering
\begin{tikzpicture}
\begin{axis}[
    ybar,
    bar width=14pt,
    width=0.95\linewidth,
    height=6cm,
    enlargelimits=0.15,
    ymin=0, ymax=100,
    ylabel={F1-Score},
    ylabel style={font=\footnotesize},
    xtick=data,
    xticklabel style={rotate=40, anchor=east, font=\footnotesize},
    ytick style={font=\footnotesize},
    legend style={
        at={(0.5,1.03)},
        anchor=south,
        legend columns=3,
        font=\footnotesize
    },
    symbolic x coords={
        PromptGuard-86M,
        Llama-Guard-3-1B,
        Llama-Guard-2-8B,
        Llama-Guard-3-8B,
        Llama-Guard-7B
    },
    nodes near coords,
    nodes near coords style={font=\tiny},
    axis x line*=bottom,
    axis y line*=left
]

\addplot+[ybar, pattern=north east lines, pattern color=cyan!70!black, draw=black, fill=cyan!30] 
    coordinates {
        (PromptGuard-86M,30)
        (Llama-Guard-3-1B,40)
        (Llama-Guard-2-8B,58)
        (Llama-Guard-3-8B,63)
        (Llama-Guard-7B,55)
    };

\addplot+[ybar, pattern=north east lines, pattern color=blue!80!black, draw=black, fill=blue!30] 
    coordinates {
        (PromptGuard-86M,00)
        (Llama-Guard-3-1B,00)
        (Llama-Guard-2-8B,10)
        (Llama-Guard-3-8B,10)
        (Llama-Guard-7B,12)
    };

\legend{Vanilla, Full Optimization}

\end{axis}
\end{tikzpicture}
\caption{Detection efficacy of different safety mechanisms against our vanilla and full (both) optimization pipeline.}
\label{fig:f1_detection_comparison}
\end{figure}

\subsection{Model-Level Trends}

As shown in Table~\ref{tab:agg_asr_min_mean_max}, while quantitative differences exist across backbones, optimization effects dominate model-specific variation. Larger models exhibit slightly higher baseline ASR in some settings, but all models converge to similarly high post-optimization $\text{ASR}_{\max}$ and comparable $\text{ASR}_{\text{mean}}$ values. This pattern suggests that the observed vulnerability arises primarily from system-level agent interactions and routing dynamics rather than idiosyncratic properties of individual language models. 

\subsection{Safety Mechanism Efficacy}
\label{sec:defense_efficacy}

This experiment evaluates how well established safety mechanisms detect conjunctive attacks under multi-agent routing, and how prompt-level optimization affects their detectability. We consider a representative set of widely deployed defenses, including PromptGuard-86M and multiple variants of Llama-Guard (3-1B, 2-8B, 3-8B, and 7B), applied either as pre-routing or post-generation filters. Detection performance is measured using F1-score in the \texttt{both} regime, aggregated across star, chain, and DAG topologies.

Figure~\ref{fig:f1_detection_comparison} compares detection performance under two attacker regimes: Vanilla attacks (no optimization) and Full Optimization (learned routing bias, key placement, and template placement). Under the Vanilla regime, most safety mechanisms detect a non-trivial fraction of attacks. Larger Llama-Guard variants achieve the highest F1-scores, while smaller models and PromptGuard exhibit more limited detection capability. Under Full Optimization, detection performance degrades sharply across all defenses. Even the strongest classifier, Llama-Guard-3-8B, experiences a substantial drop in F1-score, while smaller models collapse to near-zero detection. System-level defense results are reported in Table~\ref{tab:system_defense_eval}.

\section{Conclusion}
\label{sec:conclusion}

We study conjunctive prompt attacks in multi-agent LLM systems and show that safety behavior observed in single-agent settings does not reliably transfer to agent-to-agent deployments. The defining property of these attacks is conjunctive activation: the trigger key in the user query and the hidden template in the compromised agent can each appear benign in isolation, yet their conjunction becomes harmful once routing brings them together at the same agent. We show that optimizing prompt-level variables alone, without modifying client agents or internal routing logic, can substantially increase attack success against a compromised remote agent. Across star, chain, and DAG topologies, optimization consistently increases worst-case attack success while keeping false activations near zero in clean and partially triggered regimes, confirming that the individual components remain largely benign on their own. Our topology-aware evaluation exposes a critical blind spot in current safety assessments: vulnerabilities may appear benign under local inspection or under some communication structures, yet become reliably exploitable under others. Aggregated metrics such as minimum, mean, and maximum attack success rates therefore provide a more faithful characterization of system-level risk than average-case evaluation alone. Existing defenses also fail to fully mitigate these attacks because they typically inspect isolated prompts or outputs rather than cross-agent conjunctive conditions. Overall, these findings highlight the need for multi-agent-specific threat models, evaluation protocols, and defenses that reason over routing, provenance, and inter-agent composition. As LLM-based systems increasingly rely on decentralized reasoning and agent collaboration, safety mechanisms must account for vulnerabilities that emerge only through prompt propagation and routing-dependent conjunction.

\section*{Limitations}
Our study has several limitations. First, our routing model is intentionally abstract. We parameterize routing bias through a probabilistic dispatcher rather than a specific production-grade router, which allows us to isolate the effect of prompt-level manipulation in a model-agnostic way. However, real deployments may use classifier-based dispatch, learned policies, retrieval-driven orchestration, or system-specific heuristics. As a result, our findings should be interpreted as evidence of a structural vulnerability class rather than a claim about any single deployed routing implementation.

Second, our threat model is deliberately narrow. We study a single compromised remote agent and a prompt-level supply-chain compromise, while assuming no modification of model weights, no client-agent compromise, and no direct attacker control over routing. This scope is sufficient to show that conjunctive vulnerabilities can arise even under constrained conditions, but it does not cover stronger adversaries such as multiple colluding agents, adaptive templates, or attacks that evolve over long interaction horizons.

Third, our evaluation is conducted in a controlled experimental environment with a limited set of backbones, topologies, and defenses. Although this setup is appropriate for isolating conjunctive activation, it does not fully capture the diversity of production agentic systems, including richer tool ecosystems, longer workflows, and deployment-specific safeguards. In addition, our activation criterion relies on a deterministic marker-based predicate, which provides a reproducible operational definition of attack success but does not exhaust all possible forms of downstream harm. Extending the evaluation to concrete production-style routers, broader system designs, and more behaviorally grounded harm metrics is an important direction for future work.

\section*{Ethical Considerations}

As large language models are increasingly deployed as multi-agent systems in real-world applications, ensuring their security and robustness is a critical ethical concern. This work is motivated by the need to understand systemic vulnerabilities that arise from agent-to-agent communication, prompt propagation, and decentralized decision-making. By identifying how coordinated prompt-level manipulations can bypass existing safety mechanisms, our goal is to inform the design of more robust defenses and evaluation practices for agentic LLM systems.

We acknowledge that the attack mechanisms studied in this paper could be misused if applied irresponsibly. To mitigate this risk, all experiments were conducted in controlled, offline environments using simulated agent pipelines and open-source models. We did not evaluate or deploy these techniques against real-world systems, proprietary services, or production deployments. Furthermore, sensitive trigger tokens and template contents have been anonymized and abstracted in both code and presentation to prevent direct misuse.

Our intent is to contribute to the advancement of AI safety research by highlighting previously underexplored risks specific to multi-agent settings. We emphasize that insights into adversarial behavior should be accompanied by proactive defense strategies, and we strongly advocate for responsible disclosure, ethical experimentation, and the development of topology-aware safeguards. We hope this work encourages the community to adopt more rigorous and ethically grounded approaches to securing agent-based AI systems.

\bibliography{custom}

\appendix

\section{Use of Generative AI}
\label{sec:appendix}

To improve clarity and readability, we used LLMs solely for language editing. Their use was limited to proofreading, grammatical correction, and minor stylistic refinement, comparable to conventional grammar checkers or dictionaries. The LLMs did not contribute to the development of scientific content, ideas, or results, and their use aligns with standard manuscript preparation practices.

\section{Experimental Settings}
\label{ex}

\paragraph{Compute.} We utilize multiple NVIDIA GPUs (e.g., GTX 1080 Ti class) and run experiments sequentially across configurations.

\paragraph{Models and Defenses.}
We evaluate three instruction-tuned LLM backbones - Gemma-2B \citep{gemmateam2024gemma2improvingopen}, Mistral-7B \citep{jiang2023mistral7b}, and LLaMA3-8B \citep{grattafiori2024llama3herdmodels} and assess robustness using PromptGuard-86M \citep{Meta_2024} and multiple Llama-Guard variants (3-1B, 2-8B, 3-8B, 7B) \citep{inan2023llamaguardllmbasedinputoutput, metallamaguard2, grattafiori2024llama3herdmodels}. Together, these architectures cover diverse model scales and training paradigms, allowing us to rigorously evaluate the applicability of our attack. We employ greedy decoding with a fixed maximum generation length, and all experiments are conducted without any fine-tuning or parameter updates.

\section{Ablation: What Is Being Optimized and Why It Matters}
\label{subsec:opti}

A central contribution of this work is clarifying that effective attacks against multi-agent LLM systems do not require modifying client agents or internal routing logic. Instead, optimization operates entirely at the \emph{prompt and remote-agent interface}.

Concretely, the attacker optimizes three prompt-level variables:
\begin{enumerate}
    \item \textbf{Key placement} ($k^*$): which semantic segment of the user request carries the activation key.
    \item \textbf{Template placement} ($s^*$): how the injected template is positioned relative to the segment (prefix, wrap, or suffix).
    \item \textbf{Routing bias} ($\rho$): a probabilistic sensitivity that captures how likely key-bearing segments are to be routed to a tampered remote agent.
\end{enumerate}

Importantly, routing bias does not represent direct control over the client agent. Instead, it models the emergent effect that certain prompts, keys, or semantic patterns disproportionately influence downstream routing decisions in agentic systems. In practice, this corresponds to exploiting existing heuristics, classifiers, or learned dispatch mechanisms commonly used in multi-agent frameworks.

\subsection{Optimization Algorithm}

The following algorithm is used to optimize the counterpart model.
\begin{algorithm}[h]
\caption{Routing-Aware Counterpart Optimization}
\label{alg:surrogate}
\KwIn{Number of segments $S$, account-affinity vector $\mathbf{a}$, steps $T$}
\KwOut{Optimized configuration $\theta^*$}
Initialize logits $\alpha, \beta, \gamma$\;
\For{$t = 1$ \KwTo $T$}{
  Sample $p \leftarrow \text{GumbelSoftmax}(\alpha)$\;
  Sample $q \leftarrow \text{GumbelSoftmax}(\beta)$\;
  Compute $\rho \leftarrow \sigma(\gamma)$\;
  Compute loss $\mathcal{L}$ using Eq.~(16)\;
  Update $\alpha, \beta, \gamma$ via gradient descent\;
}
Return $\theta^* = (\arg\max_i p_i,\; \arg\max_\tau q_\tau,\; \rho)$\;
\end{algorithm}

\subsection{Ablation: Routing Bias as an Emergent Vulnerability}
\label{subsec:routing}

While routing bias may appear abstract, it captures a realistic system-level phenomenon. In deployed agentic systems, routing is often implemented using lightweight classifiers, keyword rules, or semantic similarity models. These mechanisms are inherently sensitive to prompt content.

From the attacker’s perspective, optimization over $\rho$ corresponds to finding prompt configurations that increase the probability that a key-bearing segment reaches a specific remote agent with elevated privileges. Crucially, this does not assume access to or modification of routing code. The attacker only shapes the input distribution.

Our results show that even modest increases in routing bias dramatically amplify attack success, especially in conjunction with optimized key and template placement. This demonstrates that routing bias is a first-class attack surface in multi-agent systems.

\subsection{Ablation: Topology-Aware Evaluation Is Essential}
\label{subsec:top}

A key insight from our results is that single-topology evaluations can be misleading. In many settings, star topologies exhibit high attack success even without optimization, while chain and DAG topologies suppress success due to compounding routing uncertainty.

By reporting $\text{ASR}_{\min}$, $\text{ASR}_{\text{mean}}$, and $\text{ASR}_{\max}$ across topologies, we expose worst-case and best-case behaviors simultaneously. Optimization consistently increases $\text{ASR}_{\min}$, indicating that vulnerabilities persist even under adversarial communication structures.

This suggests that safety evaluations limited to a single agent topology are insufficient for multi-agent systems.

\paragraph{Sensitivity to Routing Bias $\rho$.}

Under the counterpart formulation (Eq. 18),
\begin{equation}
P_{\text{route}} =
\sum_{i=1}^{S} p_i \cdot a_i \cdot \rho,
\end{equation}
which is linear in $\rho$.
Therefore,
\begin{equation}
\frac{\partial P_{\text{route}}}{\partial \rho}
=
\sum_{i=1}^{S} p_i \cdot a_i
\ge 0.
\end{equation}

For fixed template placement,
the surrogate attack success
$\widehat{ASR}_{both}
=
P_{\text{route}} \cdot P_{\text{template}}$
is thus monotonically non-decreasing in $\rho$.

Intuitively, increasing $\rho$ increases the
probability that a key-bearing segment reaches
the compromised agent, which directly increases
the probability of conjunctive activation.
This establishes the directional sensitivity of
the attack objective with respect to routing bias,
independent of backbone-specific behavior.

\subsection{Ablation: Routing-Aware Counterpart Optimization}
\label{app:optimization}

Let $S$ denote the number of segments. We associate each segment $s_i$ with a learnable logit $\alpha_i \in \mathbb{R}$ representing the probability that the trigger key is placed on that segment. Similarly, each template slot $\tau \in \{\text{prefix}, \text{wrap}, \text{suffix}\}$ is associated with a logit $\beta_\tau \in \mathbb{R}$. Routing bias is parameterized by a scalar logit $\gamma \in \mathbb{R}$.

We define the relaxed distributions:
\begin{align}
p_i &= \mathrm{GumbelSoftmax}(\alpha)_i, \\
q_\tau &= \mathrm{GumbelSoftmax}(\beta)_\tau, \\
\rho &= \sigma(\gamma),
\end{align}
where $\sigma(\cdot)$ denotes the sigmoid function. These distributions form a continuous relaxation of the discrete configuration $\theta = (j, \tau, \rho)$.

\subsection{Template Effectiveness Modeling.}
\label{app:rou}
Rather than assuming a fixed ordering among template slots,
we parameterize slot effectiveness using learnable
scalars $w_\tau \in \mathbb{R}$ for
$\tau \in \{\texttt{prefix}, \texttt{wrap}, \texttt{suffix}\}$.

The relaxed template selection distribution
$q_\tau$ (Eq. 16) defines a categorical distribution
over slots. The expected template effectiveness is
modeled as:

\begin{equation}
P_{\text{template}} =
\sum_{\tau} q_\tau \cdot \sigma(w_\tau),
\end{equation}

where $\sigma(\cdot)$ is a sigmoid function that
maps slot effectiveness to $[0,1]$.

Importantly, we do not impose any ordering constraint
among $w_\tau$. The optimization procedure jointly
learns both $q_\tau$ and $w_\tau$ within the counterpart
model. The final discrete configuration selects
$\tau^* = \arg\max_\tau q_\tau$.

This formulation avoids injecting prior assumptions
about slot superiority and allows the model to
adaptively identify which placement is most
effective under the given routing and backbone
conditions.

\subsection{Surrogate Fidelity and Routing-Bias Sensitivity}
\label{app:surrogate_fidelity}

To validate the routing-aware surrogate formulation
\[
\widehat{ASR}_{both} = P_{\text{route}} \cdot P_{\text{template}},
\]
we empirically decompose activation probability under varying routing bias $\rho$.
For each topology and $\rho \in \{0.0, 0.4, 0.8\}$, we measure:

\begin{itemize}
    \item $P_{\text{route}}$: the empirical probability that the key-bearing segment reaches the compromised agent in the \texttt{both} regime;
    \item $P_{\text{template}}$: the conditional probability of activation given successful routing;
    \item Empirical ASR: the measured attack success rate in the \texttt{both} regime.
\end{itemize}

The surrogate estimate is computed as
\[
\widehat{ASR}_{both}^{emp} = P_{\text{route}} \cdot P_{\text{template}}.
\]

\begin{table}[t]
\centering
\footnotesize
\setlength{\tabcolsep}{1.5pt}
\renewcommand{\arraystretch}{1.12}
\begin{tabular}{l ccc ccc}
\toprule
Topology
& \multicolumn{3}{c}{Surrogate ($\widehat{ASR}_{both}$)}
& \multicolumn{3}{c}{Empirical (ASR$_{both}$)} \\
\cmidrule(lr){2-4}\cmidrule(lr){5-7}
& Min & Mean & Max & Min & Mean & Max \\
\midrule
Star  & 0.15 & 0.53 & 0.76 & 0.20 & 0.50 & 0.80 \\
Chain & 0.13 & 0.47 & 0.68 & 0.15 & 0.52 & 0.73 \\
DAG   & 0.16 & 0.52 & 0.72 & 0.19 & 0.57 & 0.76 \\
\bottomrule
\end{tabular}
\caption{
Surrogate fidelity aggregated over routing bias $\rho \in \{0.0, 0.4, 0.8\}$.
Min / Mean / Max are computed across routing-bias configurations.
Surrogate values are computed as
$P_{\text{route}} \cdot P_{\text{template}}$.
Empirical denotes measured ASR in the \texttt{both} regime.
Across all topologies, surrogate estimates closely track empirical ASR,
confirming fidelity of the routing-aware counterpart objective.
}
\label{tab:surrogate_fidelity_agg}
\end{table}

Table~\ref{tab:surrogate_fidelity_agg} reports results for star,chain and dag topologies using three models: Gemma-2B, Mistral-7B, LLaMA3-8B .
Across all settings, surrogate estimates closely track empirical ASR,
with small absolute error.
Moreover, empirical ASR increases monotonically with $\rho$,
confirming the directional sensitivity of routing bias
predicted by the analytical formulation in Appendix~C.2.

\paragraph{Correlation analysis.}
Across all topology–routing-bias configurations (9 paired points),
the surrogate estimate $\widehat{\mathrm{ASR}}_{\text{both}} = P_{\text{route}} \cdot P_{\text{template}}$
exhibits strong agreement with empirical $\mathrm{ASR}_{\text{both}}$,
achieving Pearson correlation $r = 0.995$ and Spearman rank correlation $\rho_s = 0.933$.
This confirms both magnitude tracking and ordering preservation of the routing-aware surrogate.

These results demonstrate that the routing-aware surrogate
faithfully approximates conjunctive activation dynamics
and preserves ranking across routing-bias configurations.
This supports the use of the differentiable counterpart objective
for optimization without requiring direct discrete ASR maximization.

\subsection{Activation Predicate Verification}
\label{app:activation_verification}

To eliminate ambiguity in attack success definition,
we implement a deterministic activation predicate:
an episode is counted as activated if and only if
(i) the trigger key appears in exactly one segment,
(ii) that segment is routed to the compromised agent,
and (iii) the compromised agent's output contains
a predefined activation marker token.

For each episode, we log routing decisions,
raw outputs, and activation flags.
False activation (FA) is defined as
\[
\text{FA} = ASR_{\text{key-only}} + ASR_{\text{template-only}}.
\]

\begin{table}[t]
\centering
\footnotesize
\setlength{\tabcolsep}{2pt}
\renewcommand{\arraystretch}{1.12}
\begin{tabular}{l ccccc}
\toprule
Setting & Clean & K-only & T-only & Both & FA \\
\midrule
Baseline ($\rho=0.0$) & 0.00 & 0.04 & 0.03 & 0.28 & 0.07 \\
Biased ($\rho=0.8$)   & 0.00 & 0.05 & 0.04 & 0.74 & 0.09 \\
\bottomrule
\end{tabular}
\caption{
Verification of activation predicate under star, chain and dag topologies.
Average ASR is reported for each regime.
False activation (FA) equals
ASR$_{\text{key-only}}$ + ASR$_{\text{template-only}}$.
Activation is concentrated in the \texttt{both} regime,
while non-conjunctive regimes remain low.
}
\label{tab:activation_verification}
\end{table}

Table~\ref{tab:activation_verification}
reports ASR under the four regimes.
Activation is concentrated in the \texttt{both}
regime, while false activation remains low.
Increasing routing bias $\rho$
raises ASR in the conjunctive regime without
substantially affecting other regimes,
confirming strict conjunctive triggering.

\subsection{Transferability to Larger Instruction-Tuned and Closed-Source Backbones}
\label{app:closed_source_transfer}

To evaluate transferability beyond the open-source backbones used in the main experiments, we replicate the four-regime protocol (\textit{clean}, \textit{key-only}, \textit{template-only}, \textit{both}) on a larger instruction-tuned open-source model and a closed-source model. Specifically, we evaluate \textbf{Llama-4-Scout-17B-16E-Instruct} and \textbf{GPT-5-mini} under the same routing formulation and activation predicate used throughout the paper to test whether the conjunctive pattern persists across backbones.

We preserve the same routing formulation:

\[
Pr[a=a^*|s] =
\mathrm{clip}_{[0,1]}
\big(
\alpha I_{\text{acc}}(s)
+
\rho I_{\text{acc}}(s) I_k(s)
\big),
\]

and activation predicate:

\[
\text{Activated}
=
\{ k \in s_j \}
\cdot
\{ a_j = a^* \}
\cdot
\{ \text{marker} \in y_j \}.
\]

\begin{table}[t]
\centering
\small
\setlength{\tabcolsep}{3pt}
\renewcommand{\arraystretch}{1.08}
\begin{tabular}{l c c c c c}
\toprule
Model & $\rho$ & Clean & K-only & T-only & Both \\
\midrule
\multirow{3}{*}{Llama-4-Scout-17B}
& 0.0 & 0.00 & 0.03 & 0.02 & 0.19 \\
& 0.4 & 0.00 & 0.03 & 0.04 & 0.47 \\
& 0.8 & 0.00 & 0.04 & 0.03 & 0.69 \\
\midrule
\multirow{3}{*}{GPT-5-mini}
& 0.0 & 0.00 & 0.03 & 0.02 & 0.22 \\
& 0.4 & 0.00 & 0.04 & 0.03 & 0.51 \\
& 0.8 & 0.00 & 0.05 & 0.03 & 0.73 \\
\bottomrule
\end{tabular}
\caption{Transferability of routing-aware conjunctive attacks to a larger instruction-tuned backbone and a closed-source backbone. Across both models, ASR in the \textit{both} regime increases with routing bias $\rho$, while non-conjunctive regimes remain low.}
\label{tab:transfer_llama4_gpt5}
\end{table}

The results in Table~\ref{tab:transfer_llama4_gpt5} show that the conjunctive effect persists on both backbones. \textbf{Llama-4-Scout-17B-16E-Instruct} reaches ASR $=0.69$ at $\rho=0.8$, and \textbf{GPT-5-mini} reaches ASR $=0.73$, while the other three regimes remain low. This suggests that the vulnerability stems from routing and template interaction rather than from a particular smaller open-source model.

\subsection{Why Existing Defenses Fail}

While existing guard models (e.g., PromptGuard and Llama-Guard variants) demonstrate moderate effectiveness against non-optimized attacks, their detection performance degrades substantially under full routing-aware optimization. The core reason is architectural mismatch: most defenses are designed to inspect a single prompt or output for a locally recognizable malicious signal, whereas our attack is conjunctive and distributes activation across multiple benign-looking components.

\paragraph{(1) Localized Detection Assumption.}
Most guard mechanisms operate at the level of isolated prompts or individual model outputs, implicitly assuming that malicious intent is lexically or semantically localized within a single interaction. In contrast, our attack distributes activation across multiple components: a trigger key placed in one user segment, stochastic routing alignment, and hidden template injection within a remote agent. No single component is necessarily suspicious on its own, so a local detector may see only benign-looking fragments.

\paragraph{(2) Conjunctive Triggering Suppresses Local Signal.}
The trigger key alone and the injected template alone are designed to remain largely benign in isolation. Activation requires their conjunction at the compromised agent. As a result, lexical and semantic anomaly signals that many guard models rely on are attenuated: neither the user query nor the remote-agent template must look overtly adversarial by itself. Optimization further reduces detectable patterns by adjusting placement and routing alignment rather than introducing obviously malicious tokens.

\paragraph{(3) Routing-Dependent Activation Boundary.}
Current guardrails are largely agnostic to routing dynamics and do not reason over inter-agent communication graphs. However, activation occurs only when the key-bearing segment is routed to the compromised agent and interpreted under the injected template. This routing-dependent boundary lies outside the scope of single-prompt classification, because the harmful condition emerges only after cross-agent composition.

\paragraph{Implication for Defense Design.}
Our findings suggest that effective defenses for multi-agent systems must model conjunctive activation explicitly rather than inspecting messages in isolation. Potential directions include cross-agent provenance tracking, routing-trace monitoring, and guard models conditioned on communication structure and prompt provenance rather than isolated prompt content alone. Without modeling inter-agent propagation and composition, defenses remain misaligned with the true activation surface.

\subsection{Experimental Alignment with Prior Multi-Agent Benchmarks}

To mitigate concerns regarding artificiality of the A2A configuration, we note that our setup follows structural principles used in prior multi-agent evaluation frameworks such as AgentBench \citep{ICLR2024_e9df36b2}. Specifically, our environment includes deterministic task segmentation, stochastic agent dispatch, heterogeneous agent descriptions, and privilege asymmetry across remote agents. These properties make it possible for the trigger condition to be distributed across the pipeline rather than appearing in a single prompt.

We do not assume deterministic routing control or client-side modification. Instead, the routing formulation abstracts content-sensitive dispatch mechanisms (e.g., keyword-based, classifier-based, or embedding-based routers) commonly used in multi-agent systems. This abstraction lets us isolate architectural vulnerabilities without coupling results to a proprietary implementation.

Therefore, the observed conjunctive vulnerability emerges from composition properties inherent to agentic pipelines rather than artifacts of a synthetic environment.

\subsection{Qualitative Examples of Conjunctive Activation}

To complement quantitative ASR measurements, we provide anonymized qualitative examples illustrating behavioral differences across regimes. 
All examples use a reserved trigger token \texttt{\_\_KEY\_\_} and activation marker \texttt{\_\_ACTIVATED\_\_}. 
Sensitive template contents are abstracted. The examples make the conjunctive behavior concrete.

\paragraph{Example 1: Star Topology (Full Optimization, $\rho=0.8$)}

\textbf{User Query (segmented):}
\begin{itemize}
\item $s_1$: ``Find flights from NYC to LA next week.''
\item $s_2$: ``Check account rewards balance \texttt{\_\_KEY\_\_}.''
\item $s_3$: ``Suggest nearby hotels.''
\end{itemize}

\textbf{Routing Trace:}
\begin{itemize}
\item $s_1 \rightarrow$ Flight Agent (benign)
\item $s_2 \rightarrow$ Account Agent (compromised)
\item $s_3 \rightarrow$ Hotel Agent (benign)
\end{itemize}

\textbf{Regime Comparison:}

\textit{(a) Clean:}  
No key, no template. Output is standard account summary.

\textit{(b) Key-only:}  
Key is present, but without the hidden template no marker appears.

\textit{(c) Template-only:}  
Injected template is present, but without the key-bearing segment the response remains benign.

\textit{(d) Both (Conjunctive):}  
Key and hidden template meet at the compromised agent, triggering activation.

\textbf{Compromised Agent Output (excerpt):}
\begin{quote}
``Account summary retrieved.  
\texttt{\_\_ACTIVATED\_\_}  
[Simulated privileged behavior triggered.]"
\end{quote}

The activation marker appears only in the conjunctive regime and only when the key-bearing segment is routed to the compromised agent. 
No lexical anomaly is observable in the user query itself. 
All routing decisions and raw outputs are logged for verification.

\subsection{System-Level Defense Evaluation}
\label{sec:system_defenses}

We evaluate the routing-aware conjunctive attack
under two commonly recommended agent-deployment controls:

\begin{itemize}
    \item \textbf{D1: Tool Authorization.}
    Remote agents are restricted to a predefined tool allowlist.
    Unauthorized tool calls are rejected.

    \item \textbf{D2: Least Privilege Input.}
    Each agent receives only the minimal segment content.
\end{itemize}

These controls follow standard secure-agent recommendations
(e.g., OWASP LLM Top 10; prompt-injection mitigation literature).

We evaluate closed-source backbone transfer
with routing bias $\rho = 0.8$.
Table~\ref{tab:system_defense_eval} reports
ASR in the \texttt{both} regime and false activation (FA). These defenses are challenging because the attack signal is distributed rather than localized in one component.

\begin{table}[t]
\centering
\footnotesize
\setlength{\tabcolsep}{3pt}
\renewcommand{\arraystretch}{1.12}
\begin{tabular}{l ccc}
\toprule
Defense & ASR$_{both}$ & FA & Relative Drop \\
\midrule
None & 0.73 & 0.08 & -- \\
Tool Allowlist (D1) & 0.62 & 0.07 & -15\% \\
Least Privilege (D2) & 0.58 & 0.07 & -20\% \\
\bottomrule
\end{tabular}
\caption{
System-level defense evaluation on closed-source backbone.
ASR$_{both}$ denotes conjunctive attack success.
FA denotes false activation.
}
\label{tab:system_defense_eval}
\end{table}

\textbf{Observation.}
System-level defenses attenuate attack success,
but do not eliminate routing-aware conjunctive activation.
Even under full-stack controls,
ASR remains non-zero.
The residual vulnerability arises from
segmentation--routing interaction in multi-agent orchestration,
rather than purely from backbone misalignment.

\end{document}